\documentclass[conference,a4paper]{IEEEtran}
\IEEEoverridecommandlockouts
\pdfoutput=1

\usepackage{cite}
\usepackage{multirow}
\usepackage{graphicx}
\DeclareGraphicsExtensions{.pdf}
\usepackage[tight,footnotesize]{subfigure}
\usepackage{hyperref}
\usepackage{textpos}
\usepackage{tikz}
\usetikzlibrary{arrows,positioning,shapes,
backgrounds,external}

\usepackage{figNodes}

\hypersetup{pdfauthor={J\"urgen
    Maier and Andreas Steininger},pdftitle={Online Test Vector Insertion: A
    Concurrent Built-In Self-Testing (CBIST) Approach for Asynchronous Logic}}

\begin{document}

\setlength{\textfloatsep}{0.3\baselineskip plus 0.2\baselineskip minus 0.2\baselineskip}

\setlength{\floatsep}{0.4\baselineskip plus 0.2\baselineskip minus 0.3\baselineskip}

\title{Online Test Vector Insertion:\\A Concurrent Built-In
  Self-Testing (CBIST) Approach for Asynchronous Logic}

\author{\IEEEauthorblockN{J\"urgen Maier and Andreas Steininger}
\IEEEauthorblockA{Institute of Computer Engineering\\
Vienna University of Technology\\
Email: \{juergen.maier, andreas.steininger\}@tuwien.ac.at}
}


\maketitle

\begin{textblock*}{\textwidth}(0mm,189mm)%
  \footnotesize%
  \textcopyright\ 2014 IEEE.  Personal use of this material is permitted.
  Permission from IEEE must be obtained for all other uses, in any current or
  future media, including reprinting/republishing this material for advertising
  or promotional purposes, creating new collective works, for resale or
  redistribution to servers or lists, or reuse of any copyrighted component of
  this work in other works.%
\end{textblock*}%
\vspace*{-1em}

\begin{abstract}
Complementing concurrent checking with online testing is
crucial for preventing fault accumulation in fault-tolerant
systems with long mission times. While implementing a
non-intrusive online test is cumbersome in a synchronous
environment, this task becomes even more challenging in asynchronous designs. The latter receive increasing
attention, mainly due to their elastic timing behaviour;
however the issues related with their testing remain a key
obstacle for their wide adoption.

In this paper we present a novel approach for testing of
asynchronous circuits that leverages the redundancy present
in the conventional 4-phase protocol for implementing a
fully transparent and fully concurrent test procedure. 
The key idea is to use the protocol's unproductive
NULL phase for processing test vectors, thus effectively
interleaving the incoming 4-phase data stream with a test
data stream in a 2-phase fashion. We present implementation
templates for the fundamental building blocks required and
give a proof-of-concept by an example application that also
serves as a platform for evaluating the overheads of our
solution which turn out to be moderate.
\end{abstract}



\section{Introduction}

Throughout the last decades we have witnessed a tremendous shrinking in the
feature sizes of VLSI chips, paired with an increase of complexity. While,
without doubt, these trends have been the key to the rapidly increasing
performance, they also cause an increasing rate of faults per chip. In the face
of extremely high transistor counts and small critical charges it is unrealistic
to assume that a chip, once tested and put into operation, will perform its
operation without further experiencing transient faults or permanent
defects. Consequently, fault-tolerance provisions, e.g. based on concurrent
checking or replication and masking, have been devised to cope with those faults
and defects. However, all these approaches are based on assumptions about the
multiplicity of faults -- typically the single fault assumption -- and they will
fail when these are exceeded. While it is often sufficiently improbable that
multiple faults coincide, the potential of fault accumulation is sometimes
overlooked: A permanent fault that is tolerated within a fault-tolerance concept
still uses up its fault-tolerance capacity, thus making the system vulnerable to
the next fault that may occur, unless the first fault is properly removed. It
is, e.g., well understood that a TMR architecture exhibits lower reliability
than a simplex architecture, once one of the replica is affected by a permanent
fault. This becomes particularly cumbersome for systems with long mission
times. Therefore it is crucial, in addition to masking, to detect the existence
of a fault, diagnose and remove it.  The identification of faults may be
non-trivial, especially when faults in rarely used resources must be considered
that may remain undetected by concurrent checking approaches for a long
time. This is where on-line testing becomes mandatory~\cite{OLT}.

Asynchronous design is receiving increasing attention since it naturally avoids
some of the most serious problems currently faced by synchronous designs, such
as the need for low-skew clock distribution, insufficient tolerance to process,
temperature and voltage (PVT) variations, and high power dissipation. Instead of
a global clock it employs local handshaking to coordinate the activities, which
makes operation demand driven and timing much more flexible. One of the main
reasons why asynchronous design, although being around for several decades, has
still not been widely adopted is the difficulty of testing -- in the absence of
a clock that the tester can use to control the test procedure, even their
off-line test requires considerable efforts. In contrast, the approach we
propose here naturally leverages the redundancy already present in the
asynchronous 4-phase protocol for introducing test patterns into the data stream
in a transparent fashion and fully concurrent with the ongoing operation. The
key idea is to build components that present a conventional 4-phase interface to
the outside, but internally operate with a 2-phase protocol, which allows test
vectors to be inserted between any pair of regular data words, namely during the
NULL phase of the external protocol. At the component's output the results
pertaining to the regular data stream are presented to the outside, again in a
4-phase fashion, while the test results are internally conveyed to a response
analysis block.

The paper is structured as follows: After a review of related work we will
present the fundamental concepts of the considered asynchronous design styles in
Section~\ref{sec:background}.  Section~\ref{sec:approach} will be devoted to
presenting our approach in detail. A proof-of-concept implementation will be
given and evaluated in Section~\ref{sec:evaluation}. Finally we will conclude
the paper in Section~\ref{sec:conclusion}.


\section{Requirements and Related Work}
\label{sec:relatedwork}

Concurrent checking is a well researched field in dependable computing. Its key
principle is to employ some form of redundancy (hardware
replication~\cite{Duplication}, coding~\cite{hamming}, repeated execution of a
calculation, etc.) to allow checking whether the result of a computation is
correct.  While this approach works fine for transient faults, it is not
suitable for detecting permanent faults that may reside in a resource that is
not exercised by the ongoing operation. Several of these dormant faults may
accumulate over time and, once activated together, exceed the capabilities of
the checking scheme. In order to safely unveil these faults one cannot simply
rely on the ongoing operation to exercise the resources -- a test is needed here
that actively applies a well selected set of stimuli, independent of what is
seen through normal system operation.  This is another heavily researched area,
however most approaches were developed for synchronous circuits, which sometimes
leads to dissatisfying results when used on asynchronous ones.

We have argued above that actively applying test stimuli is desired and
characteristic for testing. At the same time these stimuli deliberately change
the state of the system under test, which interferes with the ongoing operation,
and hence seems to make testing and regular operation mutually exclusive.
Methods for online testing must fulfill two conditions: (value domain)
non-interference with the system state perceived by the application and (time
domain) no degradation of system performance beyond the point where deadlines
are missed. This can be achieved by either interleaving phases of test and
normal operation in a carefully controlled way, or by devising special test
methods that remain transparent for the ongoing operation~\cite{codeTest}.

The key quality criteria of an online test are
\begin{itemize}
 \item low performance penalty for the application
 \item high test coverage for a given fault model; this is
 determined by the quality and amount of test vectors
 \item low error detection latency; this is determined by the period
 required to apply the whole set of test vectors
\end{itemize}

We could not find approaches for a truly transparent test of asynchronous logic
in the literature. The available methods either interrupt the ongoing
operation~\cite{testInterrupt} or simply check the output without actively
applying test vectors~\cite{testCheck}.  An interesting combination of these two
models is called input vector monitoring in~\cite{IVMA}. Here a list of desired
test vectors is determined as a subset of all possible inputs during
operation. When one of these vectors is encountered during normal operation, the
corresponding output is checked against a known reference, and the vector marked
successful in the list. The test cycle completes as soon as all vectors in the
list have been marked. Variations of this scheme have been proposed that differ
in how strictly the sequence within the list must be kept; some even enter a
dedicated test mode to apply vectors that are still missing after a timeout.

The approach we propose here is specifically designed for
asynchronous logic. It provides a tight interleaving of test and
ongoing operation and exploits specific protocol properties to
largely eliminate performance penalties. It can be used with any
arbitrary set of test vectors, whose generation can be carried out by standard
methods from literature.


\section{Background}
\label{sec:background}

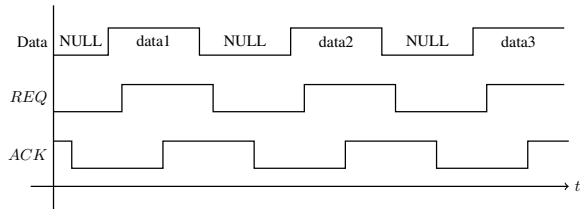
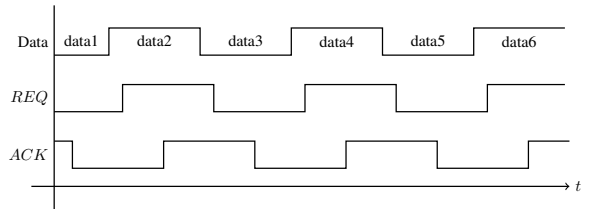
\begin{figure*}[!t]
  \centerline{
    \subfigure[4-phase protocol]{\scalebox{0.6}
      {\begin{tikzpicture}[thick]

  \draw[->] (-0.5,0) -- (11.3,0) node[right] {$t$}
  coordinate(t axis);
  \draw (0,-0.5) -- (0,4) coordinate(y axis);

  \node[left] at (0,3.2) {Data};
  \node[left] at (0,1.95) {$REQ$};
  \node[left] at (0,0.7) {$ACK$};

  \draw (0,2.9) --
  node [above, outer sep=2pt] {NULL} ++ (1.2,0)
  -- ++(0,0.6) --
  node[below, outer sep=1.2pt] {data1} ++(2,0)
  -- ++(0,-0.6) --
  node[above, outer sep=2pt] {NULL} ++(2,0)
  -- ++(0,0.6) --
  node[below, outer sep=1.5pt] {data2} ++(2,0)
  -- ++(0,-0.6) --
  node[above, outer sep=2pt] {NULL} ++(2,0)
  -- ++(0,0.6) --
  node[below, outer sep=1.5pt] {data3} ++(2,0);

  \draw (0,1.65)
  -- ++(1.5,0) -- ++(0,0.6)
  -- ++(2,0) -- ++(0,-0.6)
  -- ++(2,0) -- ++(0,0.6)
  -- ++(2,0) -- ++(0,-0.6)
  -- ++(2,0) -- ++(0,0.6)
  -- ++(1.7,0);

  \draw (0,1)
  -- ++(0.4,0) -- ++(0,-0.6)
  -- ++(2,0) -- ++(0,0.6)
  -- ++(2,0) -- ++(0,-0.6)
  -- ++(2,0) -- ++(0,0.6)
  -- ++(2,0) -- ++(0,-0.6)
  -- ++(2,0) -- ++(0,0.6)
  -- ++(0.9,0);

\end{tikzpicture}}%
      \label{fig:four-phase}}
    \hfil
    \subfigure[2-phase protocol]{\scalebox{0.6}
      {\begin{tikzpicture}[thick]

  \draw[->] (-0.5,0) -- (11.3,0) node[right] {$t$}
  coordinate(t axis);
  \draw (0,-0.5) -- (0,4) coordinate(y axis);

  \node[left] at (0,3.2) {Data};
  \node[left] at (0,1.95) {$REQ$};
  \node[left] at (0,0.7) {$ACK$};

  \draw (0,2.9) --
  node [above, outer sep=2pt] {data1} ++ (1.2,0)
  -- ++(0,0.6) --
  node[below, outer sep=1.2pt] {data2} ++(2,0)
  -- ++(0,-0.6) --
  node[above, outer sep=2pt] {data3} ++(2,0)
  -- ++(0,0.6) --
  node[below, outer sep=1.5pt] {data4} ++(2,0)
  -- ++(0,-0.6) --
  node[above, outer sep=2pt] {data5} ++(2,0)
  -- ++(0,0.6) --
  node[below, outer sep=1.5pt] {data6} ++(2,0);

  \draw (0,1.65)
  -- ++(1.5,0) -- ++(0,0.6)
  -- ++(2,0) -- ++(0,-0.6)
  -- ++(2,0) -- ++(0,0.6)
  -- ++(2,0) -- ++(0,-0.6)
  -- ++(2,0) -- ++(0,0.6)
  -- ++(1.7,0);

  \draw (0,1)
  -- ++(0.4,0) -- ++(0,-0.6)
  -- ++(2,0) -- ++(0,0.6)
  -- ++(2,0) -- ++(0,-0.6)
  -- ++(2,0) -- ++(0,0.6)
  -- ++(2,0) -- ++(0,-0.6)
  -- ++(2,0) -- ++(0,0.6)
  -- ++(0.9,0);

\end{tikzpicture}}%
      \label{fig:two-phase}}
  }
  \caption{Comparison of the two different handshake methods
  based on the indication of new data.}
  \label{fig:protocols}
\end{figure*}

In synchronous systems all activities, specifically data exchange, are
coordinated by a global clock. Asynchronous design, in contrast, employs
explicit handshaking between communicating partners~\cite{SPARSO}: The sender
indicates the validity of the data provided by means of a request ($REQ$), while
the receiver indicates their reception by means of an acknowledge ($ACK$)
signal. This closed-loop principle is the root of the elastic timing behaviour
of asynchronous designs.  Depending on the specific interpretation associated
with the transitions on $REQ$ and $ACK$ two protocols can be distinguished: In
the \emph{4-phase protocol} (see fig.~\ref{fig:four-phase}) the sender indicates
data validity by activating $REQ$, to which the receiver responds by activating
$ACK$ as soon as it has captured these data.  This is followed by a
return-to-zero phase, in which sender and receiver deactivate $REQ$ and $ACK$,
respectively.  In the \emph{2-phase protocol} (see fig.~\ref{fig:two-phase})
that unproductive return-to-zero (RTZ) phase is avoided, and the falling
transitions of $REQ$ and $ACK$ already guide the transfer of the next data
item. This halves the number of control transitions per data transfer, which
makes the 2-phase protocol the preferred choice when data needs to be
transferred in an energy-efficient way. The 4-phase protocol, on the other hand,
allows a more efficient implementation of logic functions and registers, and is
hence usually employed for computation-centric blocks.

The indication of data validity via $REQ$ faces a fundamental race condition:
The activation of $REQ$ must be perceived by the receiver only \emph{after} data
has actually become valid. The two principles used to ensure this pertain to
different timing models of the circuit and have substantially different
implementation complexity. In the bounded delay model a delay element $\Delta$
is artificially inserted into the $REQ$ signal path that is chosen large enough
to accommodate for all potential delays, including combinational functions, that
the data may experience on its travel from sender to receiver. Obviously this
necessitates a timing analysis and worst case assumptions, just like in the
synchronous case. We will further refer to this approach as \emph{bundled data
  (BD)}, since it uses one $REQ$ for the complete bundle of data. In contrast,
the delay insensitive\footnote{For the sake of simplicity we disregard the
  notion of quasi-delay insensitivity here, for a more detailed discussion see
  e.g. \cite{SPARSO}} approach uses a more elaborate coding for the data that
allows the receiver to evaluate, by means of a so-called completion detector,
when a received data item is valid.  In this way no explicit $REQ$ line is
required any more, thus avoiding the race condition. The advantage of this
solution is its ability to accommodate arbitrary delays on the data path without
the need for worst case assumptions, its drawback is the necessity of data
encoding (typically two signal rails per data bit are required). We will refer
to this approach as \emph{completion detection (CD)}. In its 4-phase version two
successive data items are separated by a so-called \emph{NULL spacer} that
establishes the RTZ phase.  In the 2-phase version the coding itself allows a
separation of successive data items.

\begin{figure}[!t]
  \centering
  \scalebox{0.6}{\begin{tikzpicture}[thick]

  \node[latchNode,draw=none] (ST_LAT) at (0,2.5) {};

  \node[lcloud] (LOG1) at (2,2.5) {comb};
  \mullerCCtrl{4,0.4}{CTR1_TOP}{CTR1_IN1}{CTR1_IN2}
  \node[latchNode] (LAT1) at (4,2.5) {latch};

  \node[lcloud] (LOG2) at (6,2.5) {comb};
  \mullerCCtrl{8,0.4}{CTR2_TOP}{CTR2_IN1}{CTR2_IN2}
  \node[latchNode] (LAT2) at (8,2.5) {latch};

  \node[lcloud] (LOG3) at (10,2.5) {comb};
  \node[latchNode,draw=none] (END_LAT) at (12,2.5) {};

  \path[->,>=stealth']
  (CTR1_TOP) edge node [right,pos=0.6] {$LE$}
  node [conn,pos=0.2] (CONN1) {} (LAT1.south)
  (CTR2_TOP) edge node [right,pos=0.6] {$LE$}
  node [conn,pos=0.2] (CONN2) {} (LAT2.south)

  (ST_LAT) edge (LOG1.west)
  (LOG1.east) edge (LAT1.west)
  (LAT1.east) edge (LOG2.west)
  (LOG2.east) edge (LAT2.west)
  (LAT2.east) edge (LOG3.west)
  (LOG3.east) edge (END_LAT);

  \node[draw] (delta) at ([xshift=1.3cm]CONN1) {$\Delta$};
  \draw (CONN1) -- (delta);

  \drawInputLeft{0.5,0.98}{REQ}{ST_CTR_TOP}
  \drawOutputLeft{0.4,-0.3}{ACK}{ST_CTR_LOW}
  \drawInputRight{11.3,0.97}{ACK}{END_CTR_TOP}
  \drawOutputRight{11.45,-0.3}{REQ}{END_CTR_LOW}

  \node[draw] (deltaIN) at ([xshift=0.9cm]ST_CTR_TOP)
       {$\Delta$};
  \draw (ST_CTR_TOP) -- (deltaIN);

  \node[draw] (deltaOUT) at ([xshift=1.3cm]CONN2)
       {$\Delta$};
  \draw (CONN2) -- (deltaOUT);

  \connCtrlForwSt{deltaIN}{CTR1_IN1}
  \connPinForw{CTR1_IN1}
  \connCtrlForw{delta}{CTR2_IN1}
  \connPinForw{CTR2_IN1}
  \connCtrlForwEnd{deltaOUT}{END_CTR_LOW.west}

  \connCtrlBackwSt{ST_CTR_LOW}{CONN1}
  \connCtrlBackw{[yshift=-0.05cm] CTR1_IN2.south}{CONN2}
  \connPinBackw{CTR1_IN2}
  \connCtrlBackwEnd{[yshift=-0.05cm] CTR2_IN2.south}
                {END_CTR_TOP.west}
  \connPinBackw{CTR2_IN2}

\end{tikzpicture}}
  \caption{Fundamental structure of an asynchronous pipeline}
  \label{fig:muller-pipe}
\end{figure}
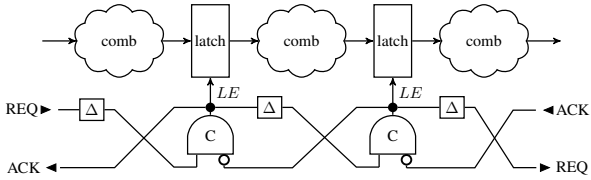

Like in the synchronous case a fundamental structure for a data processing unit
is a pipeline, in which register stages separate complex logic operations into
smaller ones. The classical pattern in the asynchronous domain is the
\emph{Muller pipeline} shown in fig.~\ref{fig:muller-pipe}. Its constituent
component is the Muller C-Element, whose function is as follows: When both
inputs match, the same value is reflected on the output; otherwise the output
retains its last value. In the 4-phase operation that we will consider in the
following, the latches in the datapath are transparent when $LE$ is active, and
opaque otherwise.

\section{Proposed Approach}
\label{sec:approach}

When comparing the data streams in fig.~\ref{fig:protocols} one can realise that
the 2-phase protocol works like a 4-phase protocol with extra data items being
conveyed during the RTZ phase. In the CD case this can be understood as
replacing the unproductive NULL spacer by productive data\footnote{For the BD
  approach we assume early data validity~\cite{SPARSO}, which is the most common
  approach anyway.}.  So when processing an incoming 4-phase data stream in a
2-phase function module, we obtain the freedom to insert a data items of our
choice in place of the NULL spacers, (ideally) without loss of performance. The
key idea of our approach is to use this freedom for inserting a stream of test
data items into the original user data stream.  Notice that, although we obtain
an extremely tight interleaving between ongoing operation and test, this
approach is completely decoupled from and transparent to the application, and
allows choosing the test data freely.

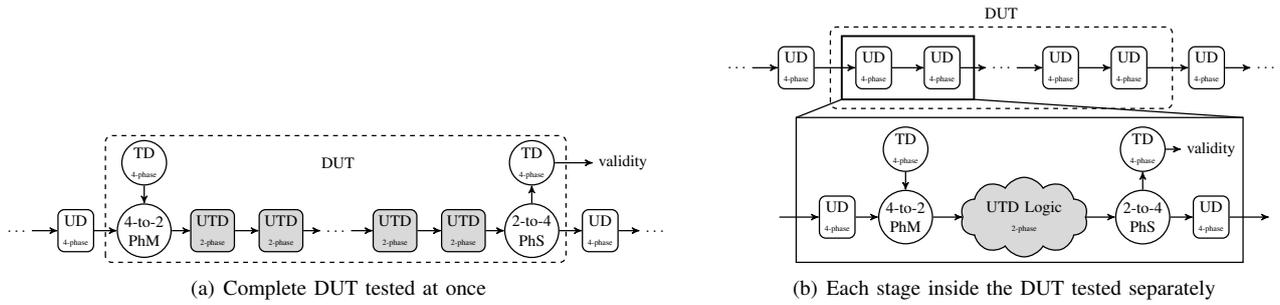
\begin{figure*}[!t]
  \centerline{
    \subfigure[Complete DUT tested at once]{\scalebox{0.6}
      {\begin{tikzpicture}[thick]

  \node (BEGIN) at (0.7,0) {$\dots$};
  \node[pipeStage] (ST1) at (2,0) {UD\\ \tiny{4-phase}};
  \node[pipeStage,pipeNew] (4TO2) at (3.5,0)
       {4-to-2 \\ PhM};
  \node[pipeStage,pipeNew] (TD1) at (3.5,1.5) {TD \\
    \tiny{4-phase}};
  \node[pipeStage,pipeAltered] (ST2) at (5,0) {UTD\\
    \tiny{2-phase}};
  \node[pipeStage,pipeAltered] (ST3) at (6.5,0) {UTD\\
    \tiny{2-phase}};
  \node (USW) at (7.7,0) {$\dots$};
  \node[pipeStage,pipeAltered] (ST4) at (9,0) {UTD\\
    \tiny{2-phase}};
  \node[pipeStage,pipeAltered] (ST5) at (10.5,0) {UTD\\
    \tiny{2-phase}};
  \node[pipeStage,pipeNew] (2TO4) at (12,0) {2-to-4 \\ PhS};
  \node[pipeStage,pipeNew] (TD2) at (12,1.5) {TD \\
    \tiny{4-phase}};
  \node[pipeStage] (ST6) at (13.5,0) {UD\\ \tiny{4-phase}};
  \node (END) at (14.7,0) {$\dots$};
  \node (VAL) at (14,1.5) {validity};

  \path[->,>=stealth',thick]
  (BEGIN) edge (ST1)
  (ST1) edge (4TO2)
  (TD1) edge (4TO2)
  (4TO2) edge (ST2)
  (ST2) edge (ST3)
  (ST3) edge (USW)
  (USW) edge (ST4)
  (ST4) edge (ST5)
  (ST5) edge (2TO4)
  (2TO4) edge (ST6)
  (2TO4) edge (TD2)
  (ST6) edge (END)
  (TD2) edge (VAL);

  \draw [rounded corners,dashed] (2.65,-0.7) rectangle
      (12.75,2.1);
  \node (CUT) at (7.75,1.5) {DUT};

\end{tikzpicture}}%
      \label{fig:CDUTT}}
    \hfil
    \subfigure[Each stage inside the DUT tested separately]
      {\scalebox{0.6}
      {\begin{tikzpicture}[node distance=3cm,thick]

  \node (BEGIN) at (0.2,0) {$\dots$};
  \node[pipeStage] (ST1) at (1.5,0) {UD\\ \tiny{4-phase}};
  \node[pipeStage] (ST2) at (3.2,0) {UD\\ \tiny{4-phase}};
  \node[pipeStage] (ST3) at (4.7,0) {UD\\ \tiny{4-phase}};
  \node (USW) at (6,0) {$\dots$};
  \node[pipeStage] (ST4) at (7.3,0) {UD\\ \tiny{4-phase}};
  \node[pipeStage] (ST5) at (8.8,0) {UD\\ \tiny{4-phase}};
  \node[pipeStage] (ST6) at (10.5,0) {UD\\ \tiny{4-phase}};
  \node (END) at (11.8,0) {$\dots$};

  \path[->,>=stealth',thick]
  (BEGIN) edge (ST1)
  (ST1) edge (ST2)
  (ST2) edge (ST3)
  (ST3) edge (USW)
  (USW) edge (ST4)
  (ST4) edge (ST5)
  (ST5) edge (ST6)
  (ST6) edge (END);

  \draw [rounded corners,dashed] (2.25,-0.9) rectangle
      (9.75,0.9);
  \draw [very thick] (2.5,-0.7) rectangle (5.4,0.7);
  \node (CUT) at (6,1.2) {DUT};
  \begin{scope}[yshift=1.7cm]

  \node (BEGIN2) at (1,-5) {};
  \node[pipeStage] (UD1) at (2.4,-5) {UD \\ \tiny{4-phase}};
  \node[pipeStage,pipeNew] (4TO2) at (3.9,-5)
       {4-to-2 \\ PhM};

  \node[pipeAltered, cloud, cloud puffs=10, draw, cloud puff
    arc=120, aspect=2, align=center, thick] (LOG) at
  (6.5,-5) {UTD Logic \\ \tiny{2-phase}};

  \node[pipeStage,pipeNew] (2TO4) at (9.1,-5)
       {2-to-4 \\ PhS};
  \node[pipeStage] (UD2) at (10.6,-5) {UD \\ \tiny{4-phase}};
  \node (END2) at (12,-5) {};
  \node[pipeStage,pipeNew] (TD1) at (3.9,-3.5)
       {TD \\ \tiny{4-phase}};
  \node[pipeStage,pipeNew] (TD2) at (9.1,-3.5)
       {TD \\ \tiny{4-phase}};
  \node (VAL) at (10.6,-3.5) {validity};
  \draw [thick] (1.5,-6) rectangle (11.3,-2.8);
  \end{scope}

  \path[->,>=stealth',thick]
  (BEGIN2) edge (UD1)
  (UD1) edge (4TO2)
  (TD1) edge (4TO2)
  (4TO2) edge (LOG)
  (LOG) edge (2TO4)
  (2TO4) edge (TD2)
  (2TO4) edge (UD2)
  (UD2) edge (END2)
  (TD2) edge (VAL);

  \draw (2.5,-0.7) -- (1.5,-1.1);
  \draw (5.4,-0.7) -- (11.3,-1.1);

\end{tikzpicture}}%
      \label{fig:SST}}
  }
  \caption{Principle of the proposed approach showing two different test
    granularities. Components that need to be adapted for the approach are
    shaded, additionally required components are shown as circles.}
  \label{fig:DUT-granularity}
\end{figure*}

Figure~\ref{fig:CDUTT} illustrates the basic architecture of our proposed
approach. At the input of the device under test (DUT) we place a \emph{4-to-2
  phase merge element} (4-to2 PhM) that joins the 4-phase user data stream
($UD$) with the 4-phase test data stream ($TD$) into a single 2-phase data
stream ($UTD$). Of course, a source for the test vectors is required here, which
is considered part of the self-testing module. In the following we will,
however, not go into detail about which test vectors to actually select, these
can be freely derived in accordance with the needs of the given DUT by means of
the available test pattern generation techniques~\cite{testing}. Here we will
only be concerned with inserting a given test vector into the data stream and
extracting the respective response later on.

The DUT now has to process the 2-phase data stream, so its design has to be
converted from the original 4-phase protocol to 2-phase. This renders it more
complex, which can somehow be considered the price for the online testing
property. The DUT's 2-phase output stream finally needs to be separated into the
test responses and the results pertaining to the application input data, which
are both again 4-phase. This task is performed by a \emph{2-to-4 phase split
  element} (2-to-4 PhS). The test responses can be analysed (compressed with a
multiple-input shift register, e.g., and compared with a stored reference)
inside the self-testing function block, while the application data stream is
passed on to the actual output where it naturally appears as the 4-phase stream
of results that one would expect in response to the original 4-phase input data
stream. So from the outside the self-testing DUT behaves like a regular 4-phase
logic block.

Interestingly, the approach allows an arbitrary choice of the DUT size: One
extreme case would be to consider every single pipeline stage a separate DUT and
equip it with all the required infrastructure at its input and output
(fig.~\ref{fig:SST}). The other extreme would be to regard the complete design
as the DUT(fig.~\ref{fig:CDUTT}), thus trading controllability and observability
for lower implementation overheads.

In the following we will focus on the description of the required merge and
split elements, since they are fundamental for our approach, and we could not
find suitable implementation patterns in the literature -- only 2-phase/4-phase
conversion of a single data stream~\cite{protConvAsync} has been considered, or
splitting and merging of datastreams following the same
protocol~\cite{protConversion,protConversionV2}.

\subsection{Merge and split for the bundled data approach}

It is possible to compose the merge unit from two nearly independent blocks, one
for handling the data bus and one for the control lines.  The data handling
block boils down to a multiplexer (MUX) that selects between forwarding the user
data and the test data. In contrast to other approaches in the
literature~\cite{protConversion} we have a strict alternation between the two
inputs and hence a fixed association between input source and state of $REQ$ in
the 2-phase protocol on the output side. This allows to hardwire the MUX's
select input to the output $REQ$, yielding low circuit complexity. In particular
we chose to associate user data with $REQ=1$ and test data with $REQ=0$.
According to the bounded delay model an appropriate delay needs to be added
before conveying the $REQ$ signal downstream, to compensate the data delay
caused by the MUX.

\begin{figure}[!t]
  \centerline{
    \subfigure[phase merge]{\scalebox{0.7}
      {\begin{tikzpicture}[scale=0.7,thick,>=to]

  \node[anchor=west] at (0.3,3.5) {C};

  \drawInputLeft{-2.5,3}{rT}{rT}
  \drawInputLeft{-2.5,4}{rU}{rU}
  \drawOutputRight{3.7,3.5}{rUT}{rUT}

  \draw (0,2.5) -- (0,4.5);
  \draw (0,2.5) .. controls (2,2.5) and (2,4.5) ..
  (0,4.5);

  \node[anchor=east,negation] (NEGC) at (0,3) {};
  \draw (rT.east) -- (NEGC);
  \draw (rU.east) -- (0,4);
  \draw (rUT.west) -- (1.5,3.5);

  \begin{scope}[yshift=0.4cm]

  \drawOutputLeft{-2.7,0}{aT}{aT}
  \drawOutputLeft{-2.7,1}{aU}{aU}
  \drawInputRight{3.5,0.5}{aUT}{aUT}

  \node[gate] (NEGB) at (0.7,0) {1};
  \node[anchor=east,negation] (NEG) at (NEGB.west) {};

  \draw (aT.east) -- (NEG);
  \draw (aUT.west) -- +(-1,0) node[conn] (CONN1) {};
  \draw (CONN1) |- (NEGB.east);
  \draw (aU.east) -| (CONN1);

  \end{scope}

\end{tikzpicture}}%
      \label{fig:phMBD}}
    \hfil
    \subfigure[phase split]
      {\scalebox{0.7}
      {\begin{tikzpicture}[scale=0.7,thick]

  \drawInputLeft{-2.5,3.5}{rUT}{rUT}
  \drawOutputRight{3.7,4}{rU}{rU}
  \drawOutputRight{3.7,3}{rT}{rT}

  \node[gate] (NEGB) at (0.25,3) {1};
  \node[negation, anchor=west] (NEG) at (NEGB.east) {};

  \draw (rT.west) -- (NEG);
  \draw (rUT.east) -- +(0.9,0)
  node[conn] (CONN1) {};
  \draw (CONN1) |- (NEGB.west);
  \draw (CONN1) |- (rU.west);

  \begin{scope}[yshift=0.4cm]

  \node[anchor=west] at (0,0.5) {C};

  \drawOutputLeft{-2.7,0.5}{aUT}{aUT}
  \drawInputRight{3.5,1}{aU}{aU}
  \drawInputRight{3.5,0}{aT}{aT}

  \draw (1,-0.5) -- (1,1.5);
  \draw (1,-0.5) .. controls (-1,-0.5) and (-1,1.5) ..
  (1,1.5);

  \node[negation, anchor=west] (NEGC) at (1,0) {};
  \draw (aT.west) -- (NEGC);
  \draw (aU.west) -- (1,1);
  \draw (aUT.east) -- (-0.5,0.5);

  \node[draw=none] at (1,-0.58) {};

  \end{scope}

\end{tikzpicture}}%
      \label{fig:phSBD}}
  }
  \caption{Handshake signals generation for the bundled data implementation}
  \label{fig:BDImpl}
\end{figure}
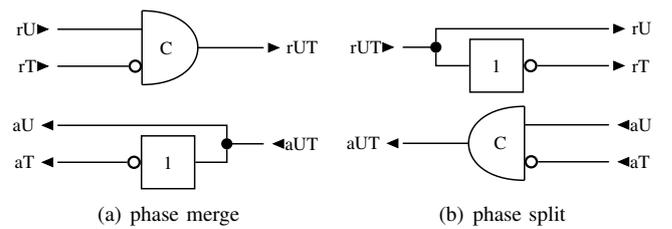

For the output $REQ$, termed $rUT$ in fig.~\ref{fig:BDImpl}, we want a rising
edge when (a) the $REQ$ of the user data ($rU$) rises, indicating new user data
are available, and (b) the $REQ$ of the test data ($rT$) falls, indicating the
test vector generator is in its RTZ phase -- whichever happens last. The same is
true for the falling edge of $rUT$, with the input transitions from $rU$ and
$rT$ inverted. The Muller C-element shown in fig.~\ref{fig:phMBD} (top) serves
exactly this purpose.  The $ACK$ signal coming from the 2-phase function unit,
termed $aUT$ can be simply conveyed as the $ACK$ to the user input (as $aU$),
and after inversion to the test input ($aT$).

For the \emph{2-to-4 phase split} element the data handling unit becomes
trivial, namely just a set of wire forks: Since the data may assume any
arbitrary value during the RTZ phase, all incoming data are directly forwarded
to both outputs at the same time. It is up to the $REQ$ signals to indicate
which of the outputs is intended to receive the respective data word.  Recall
that the merge unit associated user data with $rUT=1$. Therefore we need to
activate $rU$ (and deactivate $rT$) at the split unit output when the $rUT=1$ is
seen at its input, and set $rT=1$ (and deactivate $rU$) otherwise. The simple
circuit shown in fig.~\ref{fig:phSBD} (top) does this job and ensures that $rU$
and $rT$ are activated in a mutually exclusive fashion.  Merging the $ACK$
responses $aU$ and $aT$ from the 4-phase outputs to a common 2-phase $ACK$,
namely $aUT$, follows the same pattern as outlined for the $REQ$ signals in the
merge unit. Not surprisingly, a Muller C-element with one inverted input, as
shown in fig.~\ref{fig:phSBD} (bottom), does the job.

\subsection{Merge and split for completion detection
approach}

From the available options for implementing the CD approach we chose NCL as the
4-phase protocol and LEDR as the 2-phase one.  As these protocols use different
data representations, a bit-level conversion becomes necessary in the merge and
split unit. Table~\ref{tab:cd-conversion} shows the required mapping (per data
bit). In the 2-phase protocol we have 2 rails per bit, one value rail ($val$)
and one phase rail ($phs$). On the 4-phase side we have again 2 rails per bit,
this time a one-hot code with one rail indicating high ($hi$) and one low
($lo$). For the merge unit we need to convert from 4-phase to 2-phase
(right-to-left in the table). Notice that in the 4-phase representation only a
single rail is high at a time in each of the four valid states.

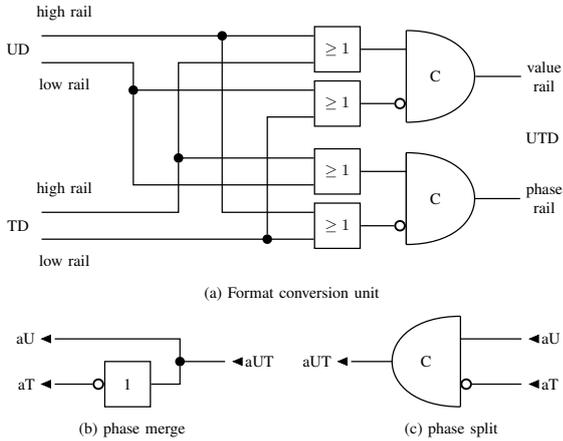
\begin{figure}[!t]
  \centering
  \scalebox{0.6}{  \begin{tikzpicture}[node distance=1.2cm, thick]

    \node[gate,draw=none] (ANDT) at (0,0) {};
    \node[gate] (OR00) at (6,0) {$\ge 1$};
    \node[gate] (OR01) [above of=OR00] {$\ge 1$};


    \node[gate] (OR10) at ([yshift=1.5cm] OR01) {$\ge 1$};
    \node[gate] (OR11) [above of=OR10] {$\ge 1$};

    \begin{scope}[node distance=3.9cm]
      \node[gate,draw=none] (ANDD) [above of=ANDT] {};
    \end{scope}

    \begin{scope}[node distance=1cm]
      \node [left of=ANDD] {UD};
      \node [left of=ANDT] {TD};
    \end{scope}

    \node[anchor= north] at ([xshift=-0.5cm]ANDT.south east)
         {low rail};

    \node[anchor= north] at ([xshift=-0.5cm]ANDD.south east)
         {low rail};

    \node[anchor= south] at ([xshift=-0.5cm]ANDT.north east)
         {high rail};

    \node[anchor= south] at ([xshift=-0.5cm]ANDD.north east)
         {high rail};

    \node[yshift=-0.6cm,xshift=0.95cm,mullerC,right of=OR11]
        (C1) {C};

    \node[yshift=-0.6cm,xshift=0.95cm,mullerC,right of=OR01]
        (C0) {C};

    \node[above right of=C0,xshift=1.5cm,yshift=0.5cm]
         {UTD};

    \draw ([xshift=-1cm]ANDD.30) -- node [near start] {} 
    +(4,0) node[conn] (I1){};
    \draw (I1) |- (OR00.150);

    \draw ([xshift=-1cm]ANDT.330) --  node [pos=0.2] {} 
    +(5,0) node[conn] (I2){};
    \draw (I2) |- (OR10.210);

    \draw (OR10.150) --  +(-4,0) node[conn] (I3){};
    \draw (I3) |- node [near end] {} 
    ([xshift=-1cm]ANDD.330);
    \draw (I3) |- (OR01.210);

    \draw (OR01.150) --  node[conn] (I4){} +(-3,0);
    \draw (I4) |- node [pos=0.83] {} 
    ([xshift=-1cm]ANDT.30);
    \draw (I4) |- (OR11.210);

    \path
    (I1) edge (OR11.150)
    (I2) edge (OR00.210);

    \draw (OR11.east) -- +(1,0);
    \node[anchor=east,negation] (NEG1) at (7.5,2.7) {};
    \draw (OR10.east) -- (NEG1);

    \draw (OR01.east) -- +(1,0);
    \node[anchor=east,negation] (NEG0) at (7.5,0) {};
    \draw (OR00.east) -- (NEG0);

    \draw (C1.north west) -- (C1.south west);
    \draw (C1.south west) .. controls
        ([xshift=2cm] C1.south west) and
        ([xshift=2cm] C1.north west) ..
        (C1.north west);

    \draw ([xshift=0.24cm] C1.east) -- node [near end] {} 
    +(1,0) node [anchor=west,align=center]
    {value\\rail};

    \draw (C0.north west) -- (C0.south west);

    \draw (C0.north west) -- (C0.south west);
    \draw (C0.south west) .. controls
        ([xshift=2cm] C0.south west) and
        ([xshift=2cm] C0.north west) ..
        (C0.north west);

    \draw ([xshift=0.22cm] C0.east) -- node [near end] {} 
    +(1,0) node [anchor=west,align=center]
    {phase\\rail};

    \node at (5,-1.5) {(a) Format conversion unit};

  \begin{scope}[yshift=-0.5cm]


  \drawOutputLeft{-0.5,-3}{aT}{aTM}
  \drawOutputLeft{-0.5,-2}{aU}{aUM}
  \drawInputRight{3.7,-2.5}{aUT}{aUTM}

  \node[gate] (NEGB) at (1.4,-3) {1};
  \node[anchor=east,negation] (NEG) at (NEGB.west) {};

  \draw (aTM.east) -- (NEG);
  \draw (aUTM.west) -- +(-1,0) node[conn] (CONN1) {};
  \draw (CONN1) |- (NEGB.east);
  \draw (aUM.east) -| (CONN1);

  \node at (1.5,-4) {(b) phase merge};


  \node[anchor=west] at (7.7,-2.5) {C};

  \drawOutputLeft{6,-2.5}{aUT}{aUTS}
  \drawInputRight{10.2,-2}{aU}{aUS}
  \drawInputRight{10.2,-3}{aT}{aTS}

  \draw (8.7,-3.5) -- (8.7,-1.5);
  \draw (8.7,-3.5) .. controls (6.7,-3.5) and (6.7,-1.5) ..
  (8.7,-1.5);

  \node[negation, anchor=west] (NEGC) at (8.7,-3) {};
  \draw (aTS.west) -- (NEGC);
  \draw (aUS.west) -- (8.7,-2);
  \draw (aUTS.east) -- (7.2,-2.5);

  \node at (8.5,-4) {(c) phase split};

  \end{scope}

  \end{tikzpicture}}
  \caption{Completion detection format conversion from
  NCL to LEDR}
  \label{fig:format-conv}
\end{figure}

The circuit shown in fig.~\ref{fig:format-conv}(a) identifies these states and
maps them to the respective LEDR code. As the two 4-phase inputs (user data and
test data) originate in different sources, we cannot avoid invalid intermediate
input patterns (i.e. such with 2 rails or no rail at high). This is why we use
Muller C-elements to retain the valid previous outputs during those phases. The
$ACK$ can be treated in the same way as in the BD merge.

In the split unit the format has to be transformed back to the 4-phase protocol
(left-to-right in Table~\ref{tab:cd-conversion}). Notice in the table how the
alternation of test data and user data in the 2-phase stream leads to a natural
insertion of the required NULL spacers into the 4-phase data streams. The
required circuit can be easily derived and is not shown here. A purely
combinational (glitch-free) implementation without Muller C-elements is
sufficient here, since the 2-phase input does not exhibit invalid intermediate
states.

Finally, the generation of the $ACK$ signal is again realised by connecting the
incoming $ACK$ lines to a Muller-C element with the test $ACK$ in its negated
form.

\begin{table}[th]
  \center
  \caption{Truth Table, Format Conversion}
  \label{tab:cd-conversion}
  \begin{tabular}{|c|c|c||c|c|c|c|c|c|}
    \hline
    \multicolumn{3}{|c||}{2-phase UTD} &
    \multicolumn{3}{c|}{4-phase UD}  &
    \multicolumn{3}{|c|}{4-phase TD} \\ \hline
    val & phs & int & hi & lo & int & hi & lo & int\\ \hline
    0 & 0 & LO(TD) & 0 & 0 & NULL & 0 & 1 & LO \\ \hline
    0 & 1 & LO(UD) & 0 & 1 & LO & 0 & 0 & NULL \\ \hline
    1 & 0 & HI(UD) & 1 & 0 & HI & 0 & 0 & NULL \\ \hline
    1 & 1 & HI(TD) & 0 & 0 & NULL & 1 & 0 & HI \\ \hline
  \end{tabular}
\end{table}

\subsection{Enhancements}
\label{sec:enhancements}

So far we have presented the basic implementations of the blocks handling the
control signals. It is possible to increase their speed at the cost of increased
complexity and thus increased area overhead. In the case of the merge element it
is possible to acknowledge the NULL phase earlier, namely as soon as the data of
the other input are propagated, giving the data values more time to travel
through the logic. This yields advantages if NULL values are much faster than
data values, which is the case when asymmetric delay lines are used.  Another
possibility is to propagate new data as soon as they show up, no matter if the
other input has already delivered its NULL spacer or not, of course only after
the $ACK$ was received from the succeeding stage. Furthermore, introducing a
latch at the output of the merge unit makes it possible to acknowledge the
inputs right away, resulting in a further decoupling of in- and output.

For the split unit it is possible, to start the NULL phase at the output that
received the last data as soon as the input gets acknowledged. Another
alternative is to acknowledge the input as soon as the output that recently
received the data has acknowledged them, without the necessity of the other one
having acknowledged its NULL phase. In addition a latch may be implemented at
the input making it possible to acknowledge the input stream right away.  For a
more detailed and generic treatment of this topic see~\cite{NNS13}.


\section{Evaluation}
\label{sec:evaluation}

We verified our online test approach for a three-stage Muller-pipeline. To keep
the focus on the newly designed units, we did not introduce combinational
functions between the pipeline registers; that would, however, be easy to add in
a next step. More specifically we augmented the pipeline by a test vector
generator, a response analyzer, and, most importantly, by our proposed merge and
split units.

\begin{figure}[!t]
  \centering
  \includegraphics[width=0.99\linewidth]{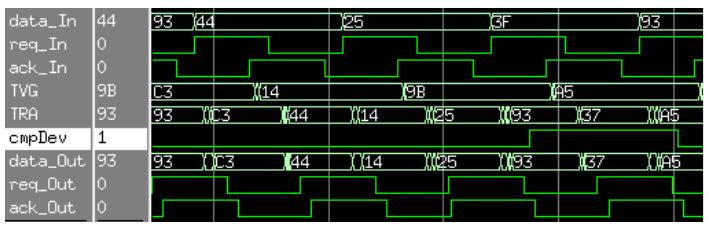}
  \caption{Post-layout simulation with detection of a stuck-at fault}
  \label{fig:eval-fault}
\end{figure}

After synthesizing the VHDL design we carried out a post-layout simulation,
whose result can be seen in fig.~\ref{fig:eval-fault}. The topmost three traces
show the input signals to the DUT and the four traces at the bottom the output
signals. $TVG$ represents the output of the test vector generator and $TRA$ the
input of the test response analyser. The highlighted signal \emph{cmpDev} gets
high as soon as $TVG$ and $TRA$ mismatch, i.e. a fault is detected.  Note how
the values from $data\_In$ and $TVG$ are processed in an alternating fashion and
show up at the output with some delay corresponding to their propagation time
through the pipeline.

To validate the self-testing capability of our approach we introduced a
stuck-at-0 fault on bit 3 in our design. This fault is activated by the test
vector $9B$ which is transformed to $93$ (as well as the data vector $3F$ being
transformed to $37$).  As soon as the TRA recognizes $93$ it raises
\emph{cmpDev}, as intended.

The area overhead and performance penalty introduced by our approach depend on
many implementation parameters and are hard to estimate in general.  We
therefore decided to give an analytic estimation here that still allows to judge
the influence of some choices, rather than presenting specific quantitative area
and timing data from the synthesized design.

For the \textit{area overhead} we compare the transistor count of the original
pipeline with that of the enhancements required for the online testing
feature. We do not include the TVG and the TRA in our analysis for two reasons:
(1) The need for these units is common to all test approaches, and (2) depending
on the specific demands the complexity of these units varies by orders of
magnitude. In general, when mapping gates to transistor counts, we did not
assume highly optimized cell designs, but we applied simplifications in the
overall circuit when they were obvious (like reducing inverter count). The
results of our analysis are shown in table~\ref{tab:eval-area}.

The first row analyses the bundled data case. In the first line the transistor
count (unit ``T'' for ``transistors'') for a stage (register plus control) of
the original pipeline is given; in case the transistor count is proportional to
the number of data bits ``/DL'' indicates ``per data line''. Line 2 gives the
\textit{overheads} for the online test. The columns correspond to the individual
function blocks (merge and split), with the rightmost column giving the overhead
in $\%$ depending on the number $n$ of stages, for large data width and without
combinational logic. For the latches the overhead for conversion into a
capture/pass latch according to \cite{Sutherland} is accounted for as
well\footnote{In contrast to the implementation proposed in \cite{Sutherland} we
  did not account $8$T per switch but rather $4$T (transmission gate).}. In the
BD approach the combinational function block (if any) remains unchanged. As
there are substantially different ways of implementing the $REQ$ delay required
in the BD approach, and furthermore the size largely varies with the required
delay value, we did not include it here. This means that the initial size of the
pipeline is underestimated here (making the relative overheads seemingly
higher), and that the extra delay to compensate for the MUX introduction is not
accounted for in the overheads.

The bottom row in table~\ref{tab:eval-area} shows the respective numbers for the
CD approach. Here the conversion of the logic is far more complicated because
each single gate has to be replaced. Unfortunately no concrete numbers could be
found in the literature: That is why we make the pessimistic assumption that the
transistor count will duplicate when moving from 4-phase to 2-phase\footnote{In
  a very simple example that we used for a first comparison, we could build an
  XOR for NCL with about $70$T, while its counterpart in LEDR required $100$T,
  yielding an overhead of less than $50\%$.}.  Furthermore, merge and split
blocks need to be added, as well as the completion detector modified.

\begin{table}[th]
  \setlength{\tabcolsep}{3pt}
  \center
  \caption{Area overhead estimated by transistor count}
  \label{tab:eval-area}
  \begin{tabular}{|c|c||c|c|c|c|c|}
    \hline
    \multicolumn{2}{|c||}{} & \textbf{merge} & \textbf{comb.} &
    \textbf{pipel. node} & \textbf{split} & \textbf{n stages (\%)}\\
    \hline \hline
    \multirow{2}{*}{\textbf{BD}} & \textbf{native} & - & - &
    $14+12/DL$ & - & \multirow{2}{*}{$117+\frac{100}{n}$}\\
    & \textbf{test}& $16+12/DL$ &  & $14/DL$ & $16$ & \\
    \hline
    \multirow{2}{*}{\textbf{CD}} & \textbf{native} & - & - &
    $2+70/DL$ & - & \multirow{2}{*}{$5.7 + \frac{94}{n}$} \\
    & \textbf{test} & $2+44/DL$ & $\approx$*$2$ & $4/DL$ & $14+22/DL$
    & \\
    \hline
  \end{tabular}
\end{table}

As shown in Table~\ref{tab:eval-area} the overhead for the BD approach is
$117+100/n\%$.  For a test-per-stage approach ($n=1$) this yields $217\%$, while
for a large number $n$ of pipeline stages between a single pair of merge/split
elements, this value drops towards $117\%$.  With large logic function blocks
this relative overhead, however, quickly approaches $0\%$: Consider a
combinational block of 12T/DL/pipeline stage; just this approximately halves the
overhead.

For the CD approach with its more complex native pipeline stages the relative
overhead is much lower.  However, as, according to our pessimistic estimation,
converting the logic function blocks roughly duplicates their transistor count,
the situation does not improve with large function blocks.

For estimating the \textit{performance penalty} we identify the additional
delays introduced by the test infrastructure. To attain a generic view we
consider gate delays (measured in inverter delays ID of the respective
technology) and assume zero wire delays. The results are summarized in
table~\ref{tab:eval-delay} ($k$ = number of data lines). They show the
accumulated values for forward and backward ($ACK$) path. In all cases we assume
that TVG and TRA operate fast enough to perform the handshaking without extra
delays.

The numbers for the BD approach are shown in the first row, with the first line
referring to the native implementation and the second one to the overhead for
the online test infrastructure. The introduction of the merge and split units
causes a delay of 5 and 2 ID, respectively. Relative to the stage delay this
represents a penalty of more than $100\%$. However, as the number of stages
between merge and split, as well as the delays of the logic function blocks (not
considered here) grow, the relative penalty quickly approaches $0\%$. Similarly,
the extra 1 ID for the register stage becomes negligible in case of complex
combinational function blocks with high delays.

For the CD approach the picture is again initially better (penalty below
$100\%$), as the native pipeline stage has more delay.  The problematic point,
however, is, once more, the complexity increase when transforming the
combinational logic from 4-phase operation to 2-phase. The related performance
penalty strongly depends on the specific circuit; we roughly estimate it as
$50...100\%$. Unfortunately, this number does not scale down with the number of
stages or with the initial complexity of the combinational logic, as in the BD
approach.

\begin{table}[th]
  \center
  \setlength{\tabcolsep}{2.5pt}
  \caption{Performance penalty in gate delays}
  \label{tab:eval-delay}
  \begin{tabular}{|c|c||c|c|c|c|c|}
    \hline
    \multicolumn{2}{|c||}{} & \textbf{merge} &
    \textbf{comb.} & \textbf{pipel. node} & \textbf{split} &
    \textbf{n stages (\%)} \\ \hline \hline
    \multirow{2}{*}{\textbf{BD}} & \textbf{native}&-  & - &
    $6$  & - &  \multirow{2}{*}{$17+\frac{117}{n}$} \\
    & \textbf{test}& $5$  & -  & $1$ & $2$  &  \\ \hline
    \multirow{2}{*}{\textbf{CD}} &\textbf{native}&-  & - &
    $9+\lceil log_2(k)\rceil *2$ & - & \multirow{2}{*}{$\frac{n+8}{n(2*\lceil log_2(k)\rceil+9)}$} \\
    & \textbf{test} & $4$ & $\approx$*$1.5-2$ & $1$ & $4$ &  \\ \hline
  \end{tabular}
\end{table}




\section{Conclusion}
\label{sec:conclusion}

We have proposed to exploit the, normally unproductive, RTZ phase or NULL
spacers of the asynchronous 4-phase protocols for conveying test vectors.  While
this can be done fully transparent and concurrent to the ongoing application, a
new test vector can be applied after every single data word, which yields the
tightest possible interleaving between test and operation, and hence an
excellent detection latency. Test vectors can be freely chosen, independent from
the user data, to optimize test coverage versus test period.  We have identified
the required infrastructure blocks for this approach and illustrated their basic
implementation. In a case study we have proven the feasibility of the approach.

For the BD approach the area overheads can, according to our estimations, go up
to $200\%$ under the most pessimistic assumptions. Fortunately they approach
$0\%$ quickly with increasing number of stages and complexity of the
combinational logic. The performance penalty can be close to $150\%$, with the
same favorable trends. So in practical cases the overheads will be very
moderate.  The CD approach exhibits lower relative overheads in the worst case
scenarios, simply because the native implementation is more complex
already. However, for the conversion of the combinational logic from 4-phase to
2-phase it is difficult to estimate the incident penalties. Under our
pessimistic assumptions the overheads for this conversion dominate, and
therefore we cannot attain the favorable scaling as seen with the BD approach.

However, even in the worst cases the observed overheads are still competitive
with those of typical fault-tolerance methods like TMR, duplication or time
redundancy, given the superior performance: In contrast to these fault-tolerance
approaches that are based on concurrent checking, our online test detects
permanent faults in the hardware, which are hard to unveil otherwise, with the
best attainable (namely cycle-wise) interleaving between application and test.

Future work will be devoted to increasing the concurrency within the merge and
split modules, as already sketched in this paper. This should aid in further
reducing the performance penalty. Furthermore, it will be interesting to study
the properties of the approach in more complex settings.





\end{document}